\documentclass[pra,twocolumn,superscriptaddress,preprintnumbers,showpacs]{revtex4}
\usepackage{amsmath,amssymb,bm,graphicx,hyperref}

\renewcommand\d{{\partial}}
\newcommand\+{\dagger}
\newcommand\<{\langle}
\renewcommand\>{\rangle}
\renewcommand\r{{\bm{r}}}
\newcommand\x{{\bm{x}}}
\renewcommand\k{{\bm{k}}}
\newcommand\kF{k_\mathrm{F}}
\newcommand\vF{v_\mathrm{F}}
\newcommand\vs{v_\mathrm{s}}
\newcommand\R{\mathrm{R}}
\renewcommand\L{\mathrm{L}}
\renewcommand\H{\mathcal{H}}

\begin{document}
\preprint{MIT-CTP 4055, INT-PUB-09-038}

\title{Universal four-component Fermi gas in one dimension}
\author{Yusuke~Nishida}
\affiliation{Center for Theoretical Physics,
Massachusetts Institute of Technology, Cambridge, Massachusetts 02139, USA}
\author{Dam~T.~Son}
\affiliation{Institute for Nuclear Theory,
University of Washington, Seattle, Washington 98195-1550, USA}

\begin{abstract}
 A four-component Fermi gas in one dimension with a short-range
 four-body interaction is shown to exhibit a one-dimensional analog of
 the BCS-BEC crossover.  Its low-energy physics is governed by a
 Tomonaga-Luttinger liquid with three spin gaps.  The spin gaps are
 exponentially small in the weak coupling (BCS) limit where they arise
 from the charge-density-wave instability, and become large in the
 strong coupling (BEC) limit because of the formation of tightly-bound
 tetramers.  We investigate the ground-state energy, the sound velocity,
 and the gap spectrum in the BCS-BEC crossover and discuss exact
 relationships valid in our system.  We also show that a one-dimensional
 analog of the Efimov effect occurs for five bosons while it is absent
 for fermions.  Our work opens up a very rich new field of universal
 few-body and many-body physics in one dimension.
\end{abstract}

\date{August 2009}


\maketitle

\section{Introduction}

Experiments using ultracold atomic gases have achieved striking progress
in realizing and studying various many-body systems previously regarded
as purely theoretical models.  One example is the Tonks-Girardeau gas in
one dimension, proposed 50 years ago~\cite{Girardeau:1960} and realized
experimentally in 2004~\cite{Paredes:2004,Kinoshita:2004}.  Another
example is the BCS-BEC crossover in a two-component Fermi gas with a
short-range two-body
interaction~\cite{Bloch:2008,Giorgini:2008,Ketterle:2008}.  It was
predicted about 40 years
ago~\cite{Eagles:1969,Leggett:1980,Nozieres:1985} and has been subject
to extensive studies after recent experimental
realization~\cite{OHara:2002,Regal:2004,Zwierlein:2004}.  In the weak
coupling limit, the system is a BCS superfluid where fermionic
excitations have an exponentially small gap, while at strong coupling it
becomes a dilute Bose-Einstein condensate of tightly-bound dimers with a
large gap for the fermionic excitations.  These two limits are smoothly
connected by varying a single parameter, the scattering length.  When
the scattering length is much larger than the range of the interaction
potential, the properties of such a system are independent of the
potential shape.  This universality makes the study of the BCS-BEC
crossover extremely worthwhile because the same properties are shared by
many different systems~\cite{Bertsch:2000}.

In this paper, we propose a purely one-dimensional analog of the BCS-BEC
crossover in a four-component Fermi gas with a short-range four-body
interaction.  The short-range four-body interaction in one dimension is
characterized by the scattering length exactly in the same way that it
characterizes the short-range two-body interaction in three
dimensions~\cite{Nishida:2008kr}, and therefore, leads to the universal
``BCS-BEC'' crossover in one dimension.  We note that while the BCS-BEC
crossover of a two-component Fermi gas in a quasi-one-dimensional
geometry has been studied before~\cite{Tokatly:2004,Fuchs:2004}, the
crossover studied in this paper has the distinction of being universal,
i.e., independent of the confinement potential.  We also note that
four-component (spin-3/2) Fermi gases with two-body interactions have
been studied and reviewed in Ref.~\cite{Wu:2006}.

In Sec.~\ref{sec:lattice}, we start with a lattice model that realizes
the BCS-BEC crossover in one dimension.  The universal regime in the
vicinity of a four-body resonance is described by a continuum theory
derived in Sec.~\ref{sec:field-theory}.  We show in
Sec.~\ref{sec:efimov} that a one-dimensional analog of the Efimov effect
occurs for five bosons, while it is absent for fermions which is
necessary for the stability of the many-body system studied in
Sec.~\ref{sec:many-body}.  We investigate the sound velocity and the gap
spectrum in the BCS limit (Sec.~\ref{sec:bcs}) and in the BEC limit
(Sec.~\ref{sec:bec}) and hypothesize that these two limits are smoothly
connected without phase transitions just as in three dimensions.  In the
unitarity limit, a one-dimensional Bertsch parameter and its connection
to the Tomonaga-Luttinger parameter are introduced in
Sec.~\ref{sec:unitarity}, whose value can be estimated in principle by
using $\epsilon$ expansions.  Exact relationships involving a contact
density are derived in Sec.~\ref{sec:exact} and the contact density is
determined from the ground state energy density in the BCS and BEC
limits.  Finally, Sec.~\ref{sec:summary} is devoted to the summary of
this paper.

\section{Few-body problems \label{sec:few-body}}

\subsection{Lattice model \label{sec:lattice}}
We start with a system of fermions with four components labeled by
$\sigma=a,b,c,d$ living on a one-dimensional lattice.  We assume that
each lattice site can accommodate one, two, or three particles with no
change in energy, but an introduction of a fourth particle into a site
with three particles releases a finite amount of energy.  The lattice
Hamiltonian for such a system is
\begin{equation}
 H = -t\!\!\sum_{\<xy\>,\,\sigma}c_{x\sigma}^\+ c_{y\sigma}
  -U\sum_x c_{xa}^\+ c_{xb}^\+ c_{xc}^\+ c_{xd}^\+
  c_{xd}c_{xc}c_{xb}c_{xa}.
\end{equation}
We will be interested in the dilute limit where the average number of
particles per site is small.  To find the universal regime, we consider
the scattering among all different components of fermions.  Such a
four-body problem is described by the Schr\"odinger equation
\begin{equation}\label{eq:schrodinger}
 \left[-t\sum_\sigma\Delta_\sigma+V(\x)\right]\Psi(\x)
  = E\,\Psi(\x),
\end{equation}
where $\x=(x_a,x_b,x_c,x_d)$ is a set of coordinates of four particles
and $\Delta_\sigma$ is the discrete Laplacian with respect to
$x_\sigma$;
$\Delta_\sigma\Psi(x_\sigma)\equiv\Psi(x_\sigma+l)+\Psi(x_\sigma-l)-2\Psi(x_\sigma)$
with $l$ being the lattice spacing.  The four-body interaction potential
is given by $V(\x)=-U$ when all $x_\sigma$ are equal and $V(\x)=0$
otherwise.

Since $V(\x)$ is translationally invariant, it is convenient to
introduce new coordinates $X=(x_a+x_b+x_c+x_d)/4$,
$r_1=(x_a+x_b-x_c-x_d)/2$, $r_2=(x_a-x_b+x_c-x_d)/2$, and
$r_3=(x_a-x_b-x_c+x_d)/2$ and assume $\Psi(\x)$ to be independent of the
center-of-mass coordinate $X$.  The Schr\"odinger
equation~(\ref{eq:schrodinger}) in terms of the remaining three relative
coordinates $\r=(r_1,r_2,r_3)$ becomes
\begin{equation}\label{eq:relative}
 \left[-t\sum_{i=1}^4\Delta_i-\delta_{\r,\bm0}U\right]\Psi(\r)
 = E\,\Psi(\r),
\end{equation}
where
$\Delta_i\Psi(\r)\equiv\Psi(\r+\bm{e}_i)+\Psi(\r-\bm{e}_i)-2\Psi(\r)$
with $\bm{e}_1=\frac{l}2(1,1,1)$, $\bm{e}_2=\frac{l}2(1,-1,-1)$,
$\bm{e}_3=\frac{l}2(-1,1,-1)$, and $\bm{e}_4=\frac{l}2(-1,-1,1)$.
Equation~(\ref{eq:relative}) is equivalent to the Schr\"odinger equation
describing one particle moving in a body-centered cubic lattice with an
attractive potential of the magnitude $U$ concentrated at one lattice
site.

One can see from Eq.~(\ref{eq:relative}) that the zero-energy wave
function at a long distance has the form
\begin{equation}\label{eq:long-distance}
 \Psi(|\r|\to\infty)\big|_{E=0} \propto \frac1{|\r|}-\frac1{a},
\end{equation}
where
$|\r|^2=\sum_{\sigma}\left(x_\sigma-X\right)^2=\frac14\sum_{\sigma<\tau}\left(x_\sigma-x_\tau\right)^2$
is the hyperradius of four particles in one dimension.  The form
(\ref{eq:long-distance}) is familiar in two-body scattering problems in
three dimensions, which can be understood from the fact that the
continuum limit of Eq.~(\ref{eq:relative}) with $l^2t\equiv\hbar^2/(2m)$
is exactly the Schr\"odinger equation in three dimensions.  Here $a$ is
an arbitrary real parameter characterizing the long-distance physics and
referred to as the scattering length.  By matching the solution of
Eq.~(\ref{eq:relative})~\cite{Watson:1939}:
\begin{equation}
 \begin{split}
  \left.\frac{\Psi(\r)}{\Psi(\bm0)}\right|_{E=0}
  &= 1-\frac{\Gamma\!\left(\frac14\right)^4}{32\pi^3}\frac{U}{t} \\
  &\quad +\frac{U}{8t}\int_{-\pi}^\pi\!\frac{d\k}{(2\pi)^3}
  \frac{e^{2i\k\cdot\r/l}}{1-\cos k_1\cos k_2\cos k_3} \\
  &\to 1-\frac{\Gamma\!\left(\frac14\right)^4}{32\pi^3}\frac{U}{t}
  +\frac{U}{8\pi t}\frac{l}{|\r|} \quad\ (|\r|\to\infty)
 \end{split}
\end{equation}
with the asymptotic form (\ref{eq:long-distance}), we find $a$ in units
of the lattice spacing $l$ to be
\begin{equation}
 \frac{l}{a}
  = \frac{\Gamma\!\left(\frac14\right)^4}{4\pi^2}-\frac{8\pi t}{U}.
\end{equation}

The scattering length $a$ can be fine-tuned to infinite corresponding to
the four-body resonance by choosing
\begin{equation}
 \frac{U}{t} =\frac{32\pi^3}{\Gamma\!\left(\frac14\right)^4}
  \approx 5.742.
\end{equation}
This value of $U/t$ separates the weak coupling regime ($a<0$) with no
bound state from the strong coupling regime ($a>0$) in which there
exists a four-body bound state (tetramer).  The wave function and
binding energy of the tetramer for $a\gg l$ are given by the universal
formulas independent of the lattice parameters:
\begin{equation}
 \Psi(|\r|\to\infty) \propto \frac{e^{-|\r|/a}}{|\r|}
  \quad\text{and}\quad E_0 = -\frac{\hbar^2}{2ma^2}.
\end{equation}
The long-distance physics near the critical value of $U/t$ should be
universal and, in particular, scale and conformal invariance are
achieved in the unitarity limit $a\to\infty$~\cite{Nishida:2008kr}.

\subsection{Field-theoretical formulation \label{sec:field-theory}}
The physics in the universal regime can be described by the following
continuum-limit Hamiltonian density (hereafter $\hbar=1$):
\begin{equation}\label{eq:hamiltonian}
 \H = - \sum_\sigma\frac{\psi_\sigma^\+\nabla^2\psi_\sigma}{2m}
  - c_0\,\psi_a^\+\psi_b^\+\psi_c^\+\psi_d^\+\psi_d\psi_c\psi_b\psi_a.
\end{equation}
Throughout this paper, we neglect two-body and three-body interactions
and interactions involving the same components of fermions.  In addition
to the translational and Galilean symmetries, the Hamiltonian density
has global U(1) and SU(4) symmetries,
\begin{equation}\label{eq:symmetry}
 \psi_\sigma\to e^{i\theta}\psi_\sigma
  \qquad\text{and}\qquad \psi_\sigma\to U_{\sigma\sigma'}\psi_{\sigma'},
\end{equation}
corresponding to the conservations of charge and SU(4) spins,
respectively.

\begin{figure}[tp]
 \includegraphics[width=\columnwidth,clip]{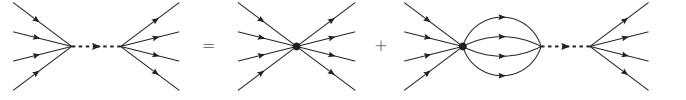}
 \caption{Feynman diagrams describing the four-body scattering in
 vacuum.  The dot represents the bare vertex $ic_0$ and the dashed line
 represents the scattering amplitude $i\mathcal{A}(E,p)$.
 \label{fig:scattering}}
\end{figure}

The second term in Eq.~(\ref{eq:hamiltonian}) describes the four-body
contact interaction among all different components of the fermionic
field $\psi_\sigma$.  $c_0$ is a cutoff-dependent coupling constant and
can be related to the above-introduced scattering length $a$ by matching
in the four-body problem.  The four-body scattering amplitude
$\mathcal{A}(E,p)$ is obtained by summing the Feynman diagrams in
Fig.~\ref{fig:scattering} into a geometric series:
\begin{equation}\label{eq:amplitude}
 \begin{split}
  &[i\mathcal{A}(E,p)]^{-1} = \frac1{ic_0} 
  + i\int_{-\infty}^\infty\frac{dk_1dk_2dk_3}{(2\pi)^3} \\
  & \times \frac{2m}
  {k_1^2+k_2^2+k_3^2+\left(k_1+k_2+k_3\right)^2+p^2/4-2mE-i0^+}.
 \end{split}
\end{equation}
Here the integrations over momenta $k_1$, $k_2$, and $k_3$ are linearly
divergent.  Introducing a momentum cutoff
$\sqrt{k_1^2+k_2^2+k_3^2}<\Lambda$ and choosing the cutoff dependence of
$c_0$ as
\begin{equation}\label{eq:coupling}
 \frac1{c_0} = \frac{m\Lambda}{3\sqrt3\pi} - \frac{m}{4\pi a},
\end{equation}
we obtain the following cutoff-independent scattering amplitude in the
limit $\Lambda\to\infty$:
\begin{equation}
 \mathcal{A}(E,p) = \frac{4\pi}{m}\frac1{-1/a+\sqrt{p^2/4-2mE-i0^+}}.
\end{equation}
In particular, when $a>0$, $\mathcal{A}(E,0)$ has a pole at a real and
negative $E$, indicating the existence of the four-body bound state.
Because its binding energy is given by $E_0=-1/(2ma^2)$, we can identify
$a$ in Eq.~(\ref{eq:coupling}) with the scattering length introduced in
Eq.~(\ref{eq:long-distance}).

\subsection{Five-body problem and Efimov effect \label{sec:efimov}}
The above arguments equally apply to four-component bosons in one
dimension.  However, many-body systems of attractive bosons tend to be
unstable to collapse, in contrast to the case of fermions where the
Pauli exclusion principle acts against such a collapse.  We now show how
such a difference already appears in a five-body problem: five bosons
develop deep bound states while five fermions do not.  This can be seen
by studying a scaling dimension of five-body composite operator
$\phi\psi_\sigma$ in the unitarity limit $a\to\infty$, where
$\phi\equiv c_0\psi_a\psi_b\psi_c\psi_d$ is a tetramer field.  If this
operator has a real scaling dimension, the corresponding five-body
system is scale invariant and thus does not support bound states.
However, if the scaling dimension is complex, the full scale invariance
is broken down to a discrete one~\cite{Nishida:2010tm}, which indicates
the formation of an infinite tower of bound states.  Such a connection
between the complex scaling dimension and the infinite tower of bound
states has been observed in resonantly-interacting three bosons or
mass-imbalanced fermions in three dimensions~\cite{Nishida:2010tm}, two
particles interacting with a $1/r^2$ potential~\cite{Moroz:2010bj}, and
the nonrelativistic AdS/CFT correspondence~\cite{Moroz:2009kv}.

\begin{figure}[tp]
 \includegraphics[width=0.9\columnwidth,clip]{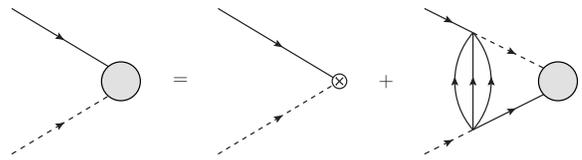}
 \caption{Feynman diagrams to renormalize five-body composite operators.
 The dashed line is a resummed propagator of $\phi$ field which is equal
 to $-i\mathcal{A}$.  The shaded bulb represents the vertex function
 $z(p)$.  \label{fig:efimov}}
\end{figure}

Feynman diagrams to renormalize $\phi\psi_\sigma$ are depicted in
Fig.~\ref{fig:efimov}.  The vertex function $z(p)$ satisfies the
following integral equation:
\begin{equation}\label{eq:vertex}
 \begin{split}
  z(p) &= 1 + \lambda\int_{-\infty}^{\infty}\!\frac{dq}{2\pi}\,
  z(q)\frac{8\pi}{\sqrt5|q|} \\
  & \times \int_{-\infty}^{\infty}\!\frac{dk_1dk_2}{(2\pi)^2}
  \frac1{\frac{2p^2+2q^2+pq}3+k_1^2+k_2^2+k_1k_2},
 \end{split}
\end{equation}
where $\lambda=\pm1$ for four-component bosons or fermions.  Because of
the scale invariance, we can assume $z(p)\propto(|p|/\Lambda)^\gamma$.
Performing the integrations in Eq.~(\ref{eq:vertex}), we find that the
anomalous dimension $\gamma$ satisfies~\cite{anomalous_dimension}
\begin{equation}\label{eq:even-parity}
 1 = -\frac{4\lambda}{\sqrt{15}}
  \frac{\cos\bigl(\gamma\arctan\!\frac1{\sqrt{15}}\bigr)}
  {\gamma\sin\bigl(\frac{\pi\gamma}2\bigr)}.
\end{equation}
The anomalous dimensions of even-parity operators satisfy the same
equation (\ref{eq:even-parity}) and their scaling dimensions are given
by
\begin{equation}\label{eq:scaling}
 \Delta_\phi+\Delta_\psi+\gamma = \frac32+\gamma.
\end{equation}

For fermions ($\lambda=-1$), we can find a series of real solutions;
$\gamma=1.59,\,4.08,\,5.99,\dots$.  According to the operator-state
correspondence~\cite{Tan:2004,Werner:2006zz,Nishida:2007pj}, each
solution corresponds to the energy of resonantly-interacting five
fermions in a one-dimensional harmonic potential by
\begin{equation}\label{eq:correspondence}
 E = \left(\frac32+\gamma\right)\omega.
\end{equation}
On the other hand, for bosons ($\lambda=+1$), in addition to real
solutions $\gamma=2.25,\,3.91,\,6.01,\dots$, we can find a pair of
complex solutions $\gamma=\pm0.735\,i$.  This is a signal of the
formation of an infinite tower of five-body bound states (pentamers)
whose spectrum exhibits the discrete scaling symmetry~\cite{spectrum}:
\begin{equation}
\frac{E_n}{E_{n+1}} = e^{2\pi/|\mathrm{Im}\gamma|} = \left(71.8\right)^2.
\end{equation}
For identical bosons with the four-body resonant interaction, the
anomalous dimension of the five-body composite operator $\phi\psi$
satisfies Eq.~(\ref{eq:even-parity}) with $\lambda=4$.  In addition to
real solutions $\gamma=3.43,\,6.02,\,8.15,\dots$, it has complex
solutions $\gamma=\pm1.25\,i$, and therefore, the spectrum of pentamers
is much denser~\cite{spectrum};
\begin{equation}
 \frac{E_n}{E_{n+1}} = e^{2\pi/|\mathrm{Im}\gamma|} = \left(12.4\right)^2.
\end{equation}
This is an analog of the Efimov effect for three identical bosons in
three dimensions~\cite{Efimov:1970}.  Note that the ordinary Efimov
effect can occur only in a spatial dimension $2.30<d<3.76$ and not in
one dimension~\cite{Nielsen:2001}.  
For comparison, the scaling factor for Efimov trimers in three
dimensions is known to be
$e^{2\pi/|\mathrm{Im}\gamma|}=\left(22.7\right)^2$.

Similarly, the anomalous dimensions of odd-parity operators [e.g.,
$4\phi(\nabla\psi_\sigma)-(\nabla\phi)\psi_\sigma$] are found to
satisfy~\cite{anomalous_dimension}
\begin{equation}\label{eq:odd-parity}
 1 = -\frac{4\lambda}{\sqrt{15}}
  \frac{\sin\bigl(\gamma\arctan\!\frac1{\sqrt{15}}\bigr)}
  {\gamma\cos\bigl(\frac{\pi\gamma}2\bigr)}
\end{equation}
with the scaling dimensions given by Eq.~(\ref{eq:scaling}).  In this
channel, both fermions and bosons have real solutions only;
$\gamma=0.833,\,3.15,\,4.87,\dots$ for $\lambda=-1$,
$\gamma=1.17,\,2.85,\,5.12,\dots$ for $\lambda=+1$, and
$\gamma=2.04,\,5.53,\,6.57,\dots$ for $\lambda=4$.  Therefore, the
corresponding states in a harmonic potential are universal and their
energies are given by
Eq.~(\ref{eq:correspondence})~\cite{Tan:2004,Werner:2006zz,Nishida:2007pj}.

\section{Many-body problems \label{sec:many-body}}

Since bosons with the four-body resonant interaction in one dimension
develop deep five-body bound states, the corresponding many-body system
cannot be stable to collapse.  Therefore, we will study the many-body
physics of four-component Fermi gas in one dimension as a function of
the dimensionless parameter $\kF a$ characterizing the short-range
four-body interaction.  Here $\kF\equiv\pi n/4$ is the Fermi momentum
defined by the total number density $n$.  In analogy with the BCS-BEC
crossover in three dimensions, we will refer to the weak (strong)
coupling limit $\kF a\to-(+)0$ as the ``BCS'' (``BEC'') limit.  We
caution that this terminology should not be taken literally, since we do
not have the spontaneous symmetry breaking in one dimension.  We shall
see below that the properties of our system in these two limits are
consistent with the crossover hypothesis that they are smoothly
connected without phase transitions.

\subsection{BCS limit \label{sec:bcs}}
The many-body physics is conveniently described by introducing the
chemical potential term $-\mu\psi_\sigma^\+\psi_\sigma$ to the
Hamiltonian density in Eq.~(\ref{eq:hamiltonian}).  In the weak coupling
(BCS) limit $\kF a\to-0$, the system develops two Fermi points
$k=\pm\kF$ and low-energy degrees of freedom are excitations around
them.  Therefore, assuming $|k\mp\kF|\ll\kF$, we can linearize the
dispersion relation and express the fermionic field in terms of two
slowly varying fields describing excitations around the Fermi points:
\begin{equation}
 \psi_\sigma(x) \simeq e^{i\kF x}\psi_\sigma^\R(x)
  + e^{-i\kF x}\psi_\sigma^\L(x).
\end{equation}
The low-energy effective theory consistent with the original symmetries
(\ref{eq:symmetry}) can be written as
\begin{equation}\label{eq:H_BCS}
 \begin{split}
  \H_\mathrm{BCS}
  &= - i\vF\,\psi_\sigma^{\R\+}\nabla\psi_\sigma^\R
  + i\vF\,\psi_\sigma^{\L\+}\nabla\psi_\sigma^\L \\
  & + g_1\,\psi_\sigma^{\L\+}\psi_\tau^{\R\+}\psi_\tau^\L\psi_\sigma^\R
  + g_2\,\psi_\sigma^{\R\+}\psi_\tau^{\L\+}\psi_\tau^\L\psi_\sigma^\R \\
  & + \frac{g_4}2\left(\psi_\sigma^{\R\+}\psi_\tau^{\R\+}
  \psi_\tau^\R\psi_\sigma^\R
  +\psi_\sigma^{\L\+}\psi_\tau^{\L\+}\psi_\tau^\L\psi_\sigma^\L\right),
 \end{split}
\end{equation}
where $\vF\equiv\kF/m$ is the Fermi velocity and summations over
$\sigma(\tau)=a,b,c,d$ are implicitly understood.  The $g_1$ term
describes the backward scattering and the $g_2$ and $g_4$ terms describe
the forward scatterings.  The low-energy parameters $g_1$, $g_2$, and
$g_4$ are determined by matching two-body scattering amplitudes at the
Fermi points with those from the microscopic theory
(\ref{eq:hamiltonian}).  To leading order in $\kF a$, we find
\begin{equation}
 g_1=g_2=g_4 = -\frac{4\vF}{\pi}\kF|a| + O\!\left[(\kF a)^2\right].
\end{equation}

The spectrum of the low-energy effective theory $\H_\mathrm{BCS}$ can be
obtained exactly by bosonization.  We introduce charge current operators
\begin{equation}
 J_0^{\R(\L)} \equiv \psi_\sigma^{\R(\L)\+}\,\psi_\sigma^{\R(\L)}
\end{equation}
and spin current operators
\begin{equation}
 J_\alpha^{\R(\L)} \equiv \psi_\sigma^{\R(\L)\+}
  \,(t_\alpha)_{\sigma\sigma'}\,\psi_{\sigma'}^{\R(\L)},
\end{equation}
where $t_\alpha$ with $\alpha=1,\dots,15$ are generators of SU(4) Lie
algebra normalized as 
$\mathrm{Tr}(t_\alpha t_\beta)=\delta_{\alpha\beta}/2$.  Using these
current operators, $\H_\mathrm{BCS}$ can be separated into two mutually
commuting parts (spin-charge
separation)~\cite{Tsvelik:1995,Assaraf:1999};
$\H_\mathrm{BCS}=\H_\mathrm{ch}+\H_\mathrm{sp}$ with
\begin{equation}\label{eq:H_charge}
 \H_\mathrm{ch} = \frac{2\pi\vF+3g_4}8
  \left(J_0^\R J_0^\R+J_0^\L J_0^\L\right) + \frac{4g_2-g_1}4J_0^\R J_0^\L
\end{equation}
and
\begin{equation}\label{eq:H_spin}
 \H_\mathrm{sp} = \sum_{\alpha=1}^{15}\left[\frac{2\pi\vF-g_4}5
  \left(J_\alpha^\R J_\alpha^\R+J_\alpha^\L J_\alpha^\L\right)
  - 2g_1 J_\alpha^\R J_\alpha^\L\right].
\end{equation}

The charge part $\H_\mathrm{ch}$ is easily diagonalized by the
Bogoliubov transformation and equivalent to the Tomonaga-Luttinger
liquid.  Introducing a bosonic field $\varphi_0(x)$ by
\begin{equation}
 \d_x\varphi_0 \equiv \frac\pi2\left(J_0^\R + J_0^\L\right)
\end{equation}
and its canonical conjugate by
\begin{equation}
 \Pi_0 \equiv -\frac12\left(J_0^\R - J_0^\L\right),
\end{equation}
the Hamiltonian density can be brought into the standard form
\begin{equation}\label{eq:tomonaga-luttinger}
 \H_\mathrm{ch} = \frac{\pi K\vs}2\Pi_0^2
  + \frac{\vs}{2\pi K}(\d_x\varphi_0)^2,
\end{equation}
which describes a gapless excitation transporting a particle number with
the linear dispersion relation $E=\pm\vs k$.  Here the
Tomonaga-Luttinger parameter $K$ and the sound velocity $\vs$ in the BCS
limit $\kF a\to-0$ are given by
\begin{equation}\label{eq:K_BCS}
 K = \sqrt{\frac{2\pi\vF+3g_4-4g_2+g_1}{2\pi\vF+3g_4+4g_2-g_1}}
  \to 1+\frac{6\kF|a|}{\pi^2}
\end{equation}
and
\begin{equation}\label{eq:vs_BCS}
 \vs = \sqrt{\left(\vF{+}\frac{3g_4}{2\pi}\right)^2
  {-}\left(\frac{4g_2{-}g_3}{2\pi}\right)^2}
  \to \vF\left(1-\frac{6\kF|a|}{\pi^2}\right).
\end{equation}
We note that the relationship $K\vs=\vF$ is guaranteed by Galilean
invariance~\cite{Cazalilla:2004}.  The Tomonaga-Luttinger parameter also
appears in other physical 
observables~\cite{Tsvelik:1995,Giamarchi:2004}, for example, in the
compressibility
\begin{equation}
 \kappa \equiv \frac{\d n}{\d\mu} = \frac{4K}{\pi\vs}
\end{equation}
and the long-distance asymptotics of correlation functions [see
Eqs.~(\ref{eq:density}) and (\ref{eq:tetramer}) below].

On the other hand, the coupling $g_1<0$ in the spin part
$\H_\mathrm{sp}$ is marginally relevant and thus opens up gaps in the
spectrum.  This can be seen by studying the renormalization group flows
of the couplings in Eq.~(\ref{eq:H_BCS}).  The straightforward one-loop
calculations result in
\begin{equation}
 \frac{dg_1}{ds}=-\frac{4g_1^2}{2\pi\vF}, \qquad
  \frac{dg_2}{ds}=-\frac{g_1^2}{2\pi\vF}, \qquad \frac{dg_4}{ds}=0,
\end{equation}
where $\Lambda_s=e^{-s}\Lambda_0$ is the momentum scale at which the
couplings $g_i(s)$ are defined.  $g_4$ and $4g_2-g_1$ are exactly 
marginal as is consistent with the fact that Eq.~(\ref{eq:H_charge}) can
be diagonalized, while $g_1$ evolves as
\begin{equation}
 g_1(s) = \frac1{\frac1{g_1(0)}+\frac{4s}{2\pi\vF}}.
\end{equation}
When $g_1(0)<0$, $g_1(s)$ reaches the Landau pole at
$s=-\frac{2\pi\vF}{4g_1(0)}=\frac{\pi^2}{8\kF|a|}$, indicating that the
second term in Eq.~(\ref{eq:H_spin}) develops spin gaps whose magnitude
is set by
\begin{equation}
 \Delta \sim \vF\Lambda_s \sim \vF\kF e^{-\pi^2/(8\kF|a|)}.
\end{equation}

The exact gap spectrum can be obtained from the Bethe-ansatz
solution if we recognize $\H_\mathrm{sp}$ as the non-Abelian part of
the SU(4) chiral Gross-Neveu
model~\cite{Andrei:1979wy,Berg:1978zn,Kurak:1978su}:
\begin{equation}\label{eq:gap_BCS}
 \Delta_f \propto
  \vF\kF e^{-\pi^2/(8\kF|a|)}\sin\!\left(\frac{f\pi}4\right).
\end{equation}
Here $f=1,2,3$ is a number of excited fermions, and accordingly, there
are three distinct gaps which are exponentially small in the BCS limit
$\kF a\to-0$.  The degeneracy of $\Delta_1$ and $\Delta_3$ can be traced
back to an accidental particle-hole symmetry in $\H_\mathrm{BCS}$ under
$\psi_\sigma^{\R(\L)}\leftrightarrow\psi_\sigma^{\R(\L)\+}$.  This
symmetry is broken by quadratic derivative terms
$\psi_\sigma^{\R(\L)\+}\nabla^2\psi_\sigma^{\R(\L)}/(2m)$ neglected in
$\H_\mathrm{BCS}$.  Because the characteristic momentum scale is set by
$\Lambda_s\sim\kF e^{-\pi^2/(8\kF|a|)}$, the small splitting of the
degeneracy is estimated to be
\begin{equation}\label{eq:gap_BCS2}
 \Delta_3-\Delta_1 \sim \frac{\Lambda_s^2}{2m}
  \sim \vF\kF e^{-\pi^2/(4\kF|a|)}.
\end{equation}

\subsection{BEC limit \label{sec:bec}}
In the strong coupling (BEC) limit $\kF a\to+0$, four fermions with all
different components form a tightly-bound tetramer and thus the
many-body system will be a dilute Bose gas of such tetramers.  In this
limit, fermionic excitations are largely gapped because of the binding
energy of the tetramer $E_0=-1/(2ma^2)$.  The gap spectrum with the
fermion number $f=1,2,3$ is simply given by
\begin{equation}
 \Delta_f \to \frac{f}{8ma^2}.
\end{equation}
Interestingly, we find that the ordering of the three spin gaps is
$\Delta_1<\Delta_2<\Delta_3$ in the BEC limit while it is
$\Delta_1\simeq\Delta_3<\Delta_2$ in the BCS limit [see
Eqs.~(\ref{eq:gap_BCS}) and (\ref{eq:gap_BCS2})].  Therefore, there has
to be a crossing between two gaps $\Delta_2$ and $\Delta_3$ as a
function of $-\infty<(\kF a)^{-1}<\infty$ in the BCS-BEC crossover.

The dilute Bose gas of tetramers to leading order in $\kF a$ is
described by the Hamiltonian density
\begin{equation}\label{eq:H_BEC}
 \H_\mathrm{BEC} = -\frac{\phi^\+\nabla^2\phi}{2M}
  - \frac1{Ma_{tt}}\phi^\+\phi^\+\phi\phi,
\end{equation}
where $M\equiv4m$ is the tetramer mass and the tetramer density is
$n_t\equiv n/4$.  $a_{tt}$ is a tetramer-tetramer scattering length
(analogous to the dimer-dimer scattering length in three
dimensions~\cite{Petrov:2004}) characterizing the scattering of two
tetramers in one dimension.  Because the scattering length $a$ is the
only scale of the system in vacuum, $a_{tt}$ has to be proportional to
$a$:
\begin{equation}
 a_{tt} = -\eta a.
\end{equation}
The coefficient $\eta$ is a universal number obtained by solving the
eight-body problem of fermions nonperturbatively.  $\eta$ is expected to
be positive because tetramers should repel each other due to the
fermionic statistics of the constituents.  If $\eta>0$, then the
many-body system of tetramers is stable.  Here we shall assume $\eta>0$
and leave the determination of the exact value of $\eta$ as a future
problem.

The effective theory of tetramers $\H_\mathrm{BEC}$ is nothing but
bosons with a $\delta$-function interaction in one dimension.  In
contrast to the dilute Bose gas in three dimensions, that in one
dimension is strongly interacting because the tetramer-tetramer coupling
in Eq.~(\ref{eq:H_BEC}) is inversely proportional to the scattering
length.  As a consequence, the tetramers in the limit $\kF a=+0$ behave
as noninteracting spinless ``fermions'' and the thermodynamic properties
of our system in the BEC limit are equivalent to those of a
noninteracting Fermi gas with the same mass $M$ and density
$n_t$~\cite{Girardeau:1960}.  Beyond such a hard-core limit, the
ground-state energy and the excitation spectrum of $\H_\mathrm{BEC}$
have been obtained exactly in Ref.~\cite{Lieb:1963rt}.  In particular,
its low-energy physics is described by the Tomonaga-Luttinger liquid
(\ref{eq:tomonaga-luttinger}) again~\cite{Cazalilla:2004}.  The
Tomonaga-Luttinger parameter $K$ and the sound velocity $\vs$ in the BEC
limit $\kF a\to+0$ are given by
\begin{equation}\label{eq:K_BEC}
 K \to 4\left(1-2n_ta_{tt}\right)
  = 4\left(1+\eta\frac{2\kF a}{\pi}\right)
\end{equation}
and
\begin{equation}\label{eq:vs_BEC}
 \vs \to \frac{\pi n_t}{M}\left(1+2n_ta_{tt}\right)
  = \frac{\vF}4\left(1-\eta\frac{2\kF a}{\pi}\right).
\end{equation}
In the expression for $K$, we have taken into account the fact that the
particle number of a tetramer is four~\cite{normalization}.  The BCS-BEC
crossover hypothesis indicates that $K$ in Eqs.~(\ref{eq:K_BCS}),
(\ref{eq:K_BEC}) and $\vs$ in Eqs.~(\ref{eq:vs_BCS}), (\ref{eq:vs_BEC})
are smoothly connected, and therefore, there has to be a maximum
(minimum) in $K$ ($\vs$) as a function of
$-\infty<(\kF a)^{-1}<\infty$.

The Tomonaga-Luttinger parameter determines the long-distance
asymptotics of correlations functions.  Because the spin degrees of
freedom are gapped in the BCS-BEC crossover, only SU(4) singlet
operators can have quasi-long-range orderings. Two such examples are the
density-density correlation function:
\begin{equation}\label{eq:density}
 \left.\<\delta n(x)\delta n(0)\>\right|_{x\to\infty}
  \to -\frac{2K}{\pi^2x^2}
  + A\frac{\cos(2\kF x)}{|\kF x|^{K/2}} + \cdots
\end{equation}
and the tetramer-tetramer correlation function:
\begin{equation}\label{eq:tetramer}
 \left.\<\phi(x)\phi^\+(0)\>\right|_{x\to\infty}
  \to \frac{B}{|\kF x|^{2/K}} + \cdots,
\end{equation}
where $A$, $B$ are unknown parameters and both $x\gg\kF^{-1}$ and
$x\gg\vF/\Delta_f$ are assumed.  We can see that the
$2\kF$-charge-density wave is the dominant order for $K\lesssim2$ (BCS
side), while the tetramer quasicondensation is the dominant order for
$K\gtrsim2$ (BEC side), and there is a crossover in between.  We note
that these correlation functions have been studied in the context of
spin-3/2 Fermi gases with two-body
interactions~\cite{Wu:2005,Lecheminant:2005,Wu:2006,Capponi:2007,Capponi:2008,Roux:2009}.

\subsection{Unitarity limit \label{sec:unitarity}}
It would be difficult to compute $K$ and $\vs$ away from the BCS or BEC
limit.  However, in the unitarity limit $\kF a\to\infty$, we can derive
exact relationships between $K$ and thermodynamic quantities.  Because
the density $n$ is the only scale of the system, the ground state energy
density of the unitary Fermi gas can be written as
\begin{equation}\label{eq:xi}
 \mathcal{E}_\mathrm{unitary}(n) \equiv \xi \mathcal{E}_\mathrm{free}(n),
\end{equation}
where the ground state energy density of a noninteracting Fermi gas is
\begin{equation}
 \mathcal{E}_\mathrm{free}(n) = \frac{\pi^2}{96m}n^3.
\end{equation}
Here $\xi$, which measures how much energy is gained due to the
attractive interaction, is a universal number to characterize the
strongly-interacting unitary Fermi gas and analogous to the Bertsch
parameter in three dimensions~\cite{Bertsch:2000}.  From the
thermodynamic relationships, we obtain the pressure as
$P(n)=2\mathcal{E}(n)$, and thus, the sound velocity is given by
\begin{equation}
 \vs^2=\frac1{m}\frac{\d P}{\d n}=\xi\vF^2.
\end{equation}
Because $K\vs=\vF$ is guaranteed by Galilean
invariance~\cite{Cazalilla:2004}, we find that the Tomonaga-Luttinger
parameter is related to the one-dimensional Bertsch parameter by
\begin{equation}
 K = \frac1{\sqrt{\xi}}.
\end{equation}
This relationship implies $K>1$ in the unitarity limit because $0<\xi<1$
is expected for the attractive interaction.  It is a challenging
many-body problem to determine the exact value of $\xi$.

One possible way to estimate the value of $\xi$ is to use the $\epsilon$
expansion~\cite{Nishida:2006br,Nishida:2010tm}.  Considering the same
Hamiltonian density (\ref{eq:hamiltonian}) in an arbitrary spatial
dimension $d$, we find that the dimension of the coupling constant $c_0$
is given by $\nu=2-3d$, which also determines the behavior of the
four-body wave function at a short distance:
\begin{equation}
 \Psi(|\r|\to0) \to |\r|^\nu.
\end{equation}
In the limit $d\to2/3$, we have $\nu\to0$ and the singularity in the
wave function disappears.  This means that the contact interaction among
four fermions disappears and thus the unitary Fermi gas reduces to a
noninteracting Fermi gas~\cite{amplitude}.  On the other hand, in the
limit $d\to4/3$, we have $\nu\to-2$ so that the normalization integral
of the wave function
\begin{equation}
 \int\,d^{3d}\r|\Psi(\r)|^2 \sim \int_0dr\,r^{3-3d}
\end{equation}
diverges at the origin $|\r|\to0$.  This means that four fermions behave
as a point-like composite boson and thus the unitary Fermi gas reduces
to a noninteracting Bose gas of such
tetramers~\cite{Nussinov:2006zz,amplitude}.  Therefore, $\xi$ defined as
in Eq.~(\ref{eq:xi}) is found to be
\begin{equation}
 \left.\xi\right|_{d\to\frac23} \to 1
  \qquad\text{and}\qquad
  \left.\xi\right|_{d\to\frac43} \to 0.
\end{equation}
It is possible to formulate appropriate perturbation theories around
these critical dimensions~\cite{epsilon}.  Interpolations of two
systematic expansions in terms of $\bar\epsilon=d-\frac23$ and
$\epsilon=\frac43-d$ would provide a reasonable estimate of $\xi$ in
$d=1$ as in three dimensions~\cite{Nishida:2006br,Nishida:2010tm}.

\subsection{Exact relationships \label{sec:exact}}
\begin{figure}[tp]
 \includegraphics[width=0.6\columnwidth,clip]{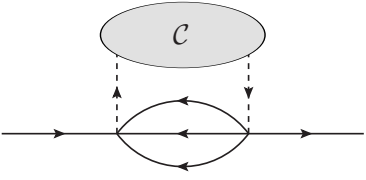}
 \caption{Contribution of the contact density $\mathcal{C}$ to the
 fermion propagator $iG_\sigma(k_0,k)$.  The integration over $k_0$
 leads to the momentum distribution in Eq.~(\ref{eq:distribution}).
 \label{fig:contact}}
\end{figure}

Another characteristic of our system (\ref{eq:hamiltonian}) that
resembles the BCS-BEC crossover in three dimensions is the
large-momentum tail of the momentum distribution of fermions and its
relationships to other properties of the system~\cite{Tan:2005}.  In
order to see this, we consider the following operator product expansion
(no sum over $\sigma=a,b,c,d$):
\begin{equation}
 \begin{split}
  & \psi_\sigma^\+\!\left(x-\frac{y}2\right)
  \psi_\sigma\!\left(x+\frac{y}2\right)
  = \psi_\sigma^\+\psi_\sigma\!(x)
  + \frac{y}2\psi_\sigma^\+\tensor{\nabla}\psi_\sigma(x) \\
  &\ - \frac{\sqrt3|y|}{8\pi}(mc_0)^2
  \psi_a^\+\psi_b^\+\psi_c^\+\psi_d^\+\psi_d\psi_c\psi_b\psi_a(x)
  + O(y^2).
 \end{split}
\end{equation}
This can be confirmed by evaluating expectation values of the both sides
for a state consisting of four fermions with all different
components~\cite{Braaten:2008uh}.  The nonanalytic term $\sim|y|$
indicates that the momentum distribution of fermions
\begin{equation}
 \rho_\sigma(k) \equiv \int\!dy\,e^{-iky}
  \left\<\psi_\sigma^\+\!\left(x-\frac{y}2\right)
   \psi_\sigma\!\left(x+\frac{y}2\right)\right\>
\end{equation}
falls off by a power of $|k|\to\infty$ as
\begin{equation}\label{eq:distribution}
 \rho_\sigma(k) \to \frac{\sqrt3}{4\pi}\frac{\mathcal{C}}{k^2}.
\end{equation}
The coefficient is given by the so-called contact density:
\begin{equation}\label{eq:contact}
 \mathcal{C} \equiv \<(mc_0)^2
  \psi_a^\+\psi_b^\+\psi_c^\+\psi_d^\+\psi_d\psi_c\psi_b\psi_a\>.
\end{equation}
The same result can be obtained by the method used in
Ref.~\cite{Son:2010kq} (see Fig.~\ref{fig:contact}).

From Eqs.~(\ref{eq:hamiltonian}), (\ref{eq:coupling}), and
(\ref{eq:contact}), we find that the energy density of the system
$\mathcal{E}\equiv\<\H\>$ can be expressed by
\begin{equation}
 \mathcal{E} = \sum_\sigma\int_{-\infty}^\infty\!\frac{dk}{2\pi}\frac{k^2}{2m}
  \left(\rho_\sigma(k)-\frac{\sqrt3}{4\pi}\frac{\mathcal{C}}{k^2}\right)
  + \frac{\mathcal{C}}{4\pi ma}.
\end{equation}
This relationship is valid for any state of the system and for any value
of the scattering length $a$.  Derivations of the pressure
\begin{equation}
 P = 2\mathcal{E} + \frac{\mathcal{C}}{4\pi ma},
\end{equation}
the adiabatic relationship
\begin{equation}\label{eq:adiabatic}
 \frac{d\mathcal{E}}{da} = \frac{\mathcal{C}}{4\pi ma^2},
\end{equation}
and the generalized virial theorem in the presence of a harmonic
potential
$V_\omega=\int\!dx\,\frac12m\omega^2x^2\sum_\sigma\psi_\sigma^\+(x)\psi_\sigma(x)$:
\begin{equation}
 E = 2\<V_\omega\> - \int\!dx\,\frac{\mathcal{C}(x)}{8\pi ma}
\end{equation}
are straightforward from the above results by using the methods in
Refs.~\cite{Tan:2005,Braaten:2008uh}.

From the adiabatic relationship (\ref{eq:adiabatic}) together with the
ground state energy density in the BCS limit (up to the mean-field
correction):
\begin{equation}
 \mathcal{E}_\mathrm{BCS} = \mathcal{E}_\mathrm{free}
  + \frac{4\pi a}{m}\left(\frac{n}4\right)^4 + O\!\left[\kF^5a^2\right]
\end{equation}
and in the BEC limit~\cite{Lieb:1963rt}:
\begin{align}
 \mathcal{E}_\mathrm{BEC} &= E_0n_t + \frac{\pi^2}{6M}n_t^3
  + \frac{\pi^2}{3M}n_t^4a_{tt} + O\!\left[\kF^5a^2\right],
\end{align}
we find that the contact density $\mathcal{C}$ is given by
\begin{equation}
 \frac{\mathcal{C}}{\kF^2} \to \frac{16\kF^2a^2}{\pi^2} \qquad (\kF a\to-0)
\end{equation}
and
\begin{equation}
 \frac{\mathcal{C}}{\kF^2} \to \frac4{\kF a} - \eta\frac{\kF^2a^2}{3\pi}
  \qquad (\kF a\to+0). 
\end{equation}
The BCS-BEC crossover hypothesis indicates that both $\mathcal{E}$ and
$\mathcal{C}$ smoothly evolve as functions of
$-\infty<(\kF a)^{-1}<\infty$.

References~\cite{Tan:2005,Braaten:2008uh} have also shown that the
contact density in three dimensions is related to the local pair
density, which is the number of pairs of spin-up and -down fermions with
small separations.  Similarly, the contact density in our
one-dimensional system is related to the local quadruplet density
$\mathcal{N}_4(R)$, which is the number of sets of four-component
fermions with small hyperradii.  This can be seen from the following
operator product expansion:
\begin{equation}
 \begin{split}
  &\psi_a^\+\psi_a(x_a)\psi_b^\+\psi_b(x_b)
  \psi_c^\+\psi_c(x_c)\psi_d^\+\psi_d(x_d) \\
  &= \frac{(mc_0)^2\psi_a^\+\psi_b^\+\psi_c^\+\psi_d^\+\psi_d\psi_c\psi_b\psi_a(X)}
  {16\pi^2\left(r_1^2+r_2^2+r_3^2\right)} + O(|\r|^{-1}).
 \end{split}
\end{equation}
The integral of the left-hand side over the three relative coordinates
$|\r|<R$ counts the number of sets of four-component fermions at the
fixed center-of-mass coordinate $X$ but with the hyperradius smaller
than $R$.  Therefore, we find that the short-distance asymptotics of the
local quadruplet density is related to the contact density by
\begin{equation}
 \mathcal{N}_4(R\to0) \to \frac{\mathcal{C}}{4\pi}R.
\end{equation}

\section{Conclusions \label{sec:summary}}

In summary, we have demonstrated that the four-component Fermi gas in
one dimension exhibits the one-dimensional analog of the BCS-BEC
crossover as a function of the scattering length characterizing the
short-range four-body interaction.  We investigated the ground-state
energy, the sound velocity, the gap spectrum, and the exact
relationships in the BCS-BEC crossover and found that the gap spectrum
has the rich structure because of the existence of three distinct gaps.
We also showed that the one-dimensional analog of the Efimov effect
occurs for five bosons while it is absent for fermions.  This work
extends our perspectives on the universal few-body and many-body physics
to one dimension and possibly opens up a very rich new research area.
Finally, we note that the system considered in this paper is highly
fine-tuned: not only the four-body interaction is tuned to the
resonance, two-body and three-body interactions have to be tuned to
vanish.  Its experimental realization would be challenging.

\acknowledgments
The authors thank Shina Tan for discussions.  Y.\,N.\ was supported by
MIT Pappalardo Fellowship in Physics and DOE Office of Nuclear Physics
under grant DE-FG02-94ER40818.  This work was supported, in part, by DOE
Grant No.\ DE-FG02-00ER41132.


\begin{thebibliography}{99}
\bibitem{spectrum}
This spectrum in the unitarity limit has been confirmed by numerically
solving the integral equation to determine the petramer's binding energy
$E<-\frac{\theta(a)}{2ma^2}$:
\begin{equation}\nonumber
 \begin{split}
  Z_\pm(p) &= \lambda\int_{-\Lambda}^{\Lambda}\!\frac{dq}{2\pi}\,
  Z_\pm(q)\frac{4\pi}{-\frac1{a}+\sqrt{\frac{5q^2}4+2m|E|}} \\
  & \times \int_{-\Lambda}^{\Lambda}\!\frac{dk_1dk_2}{(2\pi)^2}
  \frac1{\frac{2p^2+2q^2+pq}3+k_1^2+k_2^2+k_1k_2+m|E|},
 \end{split}
\end{equation}
where $Z_\pm(-p)=\pm Z_\pm(p)$ correspond to the even- and odd-parity
channels, respectively.
\bibitem{normalization}
Note that $\varphi_0(x)$ in Eq.~(\ref{eq:tomonaga-luttinger}) is
normalized as $\d_x\varphi_0=\frac\pi2\delta n(x)$, where
$n+\delta n(x)=\sum_\sigma\psi_\sigma^\+\psi_\sigma=4\phi^\+\phi$ is the
number density operator.
\bibitem{amplitude}
This observation can be supported by extending the four-body scattering
amplitude in Eq.~(\ref{eq:amplitude}) to an arbitrary spatial dimension:
\begin{equation}\nonumber
 \left.\mathcal{A}(E,\bm{p})\right|_{a=\infty}
  = -\left(\frac{2\pi}m\right)^{\frac{3d}2}
  \frac{2^d}{\Gamma\!\left(1-\frac{3d}2\right)
  \left[\frac{\bm{p}^2}{8m}-E-i0^+\right]^{\frac{3d}2-1}}.
\end{equation}
When $d\to2/3$ or $d\to4/3$, we find $\mathcal{A}\to0$ corresponding to
the noninteracting limit.
\bibitem{epsilon}
We have carried out the expansion in terms of $\bar\epsilon=d-\frac23$
up to its next-to-leading order and found
\begin{equation}\nonumber
 \left.\xi\right|_{d=\frac23+\bar\epsilon}
  = 1 - \frac{81\,\bar\epsilon}{2^{1/3}\Gamma\!\left(\frac13\right)^3}
  + O(\bar\epsilon^2). 
\end{equation}
Due to the large $O(\bar\epsilon)$ correction, the naive extrapolation
to $d\to1$ does not work ($\xi\to-0.115$) and needs to be combined with
the expansion in terms of $\epsilon=\frac43-d$.
\end{thebibliography}
\end{document}